\documentstyle[osa,aplop,manuscript]{revtex}

\begin{document}

\title{Low Energy Nucleon-Nucleon Scattering with the Skyrme Model
in the Geodetic Approximation}

\author{by T. Gisiger and M. B. Paranjape}

\address{Laboratoire de physique nucl\'eaire, Universit\'e de
Montr\'eal C.P. 6128, succ ``A", Montr\'eal, Qu\'ebec, Canada, H3C 3J7}

\maketitle

\begin{abstract}
We calculate nucleon-nucleon scattering at low energies and large impact
parameter in the Skyrme model within the framework for soliton scattering
proposed by Manton. This corresponds to a truncation of the degrees of
freedom to the twelve collective coordinates which essentially describe the
rigid body motion of the pair of Skyrmions. We take to its logical conclusion
the result that the induced kinetic energy for these collective coordinates in
the product ansatz behaves as one over the separation and hence can dominate
over the potential. This behaviour implies to leading order that we can drop
the potential and the resulting motion reduces simply to geodesic motion on the
manifold parametrized by the variables of the product ansatz. We formulate the
semi-classical quantization of these variables to obtain the motion
corresponding to the nucleonic states of the Skyrme model. This is the
appropriate description for the nucleons in order to consider their scattering
within Manton's framework in the semi-classical approximation. We investigate
the implications for the scattering of nucleons with various initial
polarizations using the approximation method of ``variation of
constants''.
\end{abstract}

\section{ Introduction}
The Skyrme model\cite{1} is unique among models describing the strong
interactions.
It is conceived from the very basic and fundamental principles concerning the
symmetries
and their realizations in the strong interaction, and it can be used to make
quantitative predictions about low energy hadronic dynamics. It contains
exactly two parameters, $f_{\pi}$ and $e$, which can in principle be computed
from QCD. In practice, they are fitted by phenomenological consideration. Once
these parameters are fixed, the model can make reasonably accurate predictions
in the meson sector\cite{2} and for the static properties of single
baryons\cite{3}.

The case $B=2$ is very interesting since it contains both the scattering of two
nucleons and their bound state, the deuteron. Much effort has been invested in
extracting the nucleon-nucleon potential\cite{4}. This comprises of a
calculation
of the potential energy of two well separated Skyrmions, taken in the product
ansatz or slight variations thereof, completed with a projection onto
asymptotic quantum nucleonic states. For large separation $d$ between the
nucleons, the static potential behaves as $1/d^3$, for the case of massless
pions, and contributes only to the spin-spin and tensor channels of the
nucleon-nucleon potential. Adding a pion mass cuts off all interactions with
the usual Yukawa decaying exponential, however the potential now has a leading
$1/d$ behaviour in the tensor channel and a $1/d^2$ central channel, albeit,
both are multiplied by two powers
of the pion mass. The product ansatz is deficient to produce an intermediate
range attraction in the central channel, however, it has been demonstrated by
exact numerical calculations unfettered by this ansatz, that the Skyrme model
does indeed contain a central attraction\cite{5,6}. The Skyrme model has not
been
too useful for determining the potential at small separation. The exact
deuteron
profile is thought to be a toroidal configuration where the Skyrmions loose
their identity. In this way it is clear that the product ansatz will fail to
properly describe the situation, since it distinctly identifies the two
Skyrmions.

It is the aim of this paper to consider the influence of the kinetic energy on
the nucleon-nucleon interaction. Within the framework of Manton\cite{7} for
soliton
scattering, the system is truncated to the finite number of degrees of freedom
parametrizing the low energy critical points of the static configuration and
the gradient flow curves joining them together. The full potential serves to
induce a potential on this sub-manifold, while the full kinetic energy serves
to induce a metric. We, and also Schroers\cite{16} have computed the induced
kinetic energy for the case of
massless pions\cite{8} and found a leading term which behaves as $1/d$. This
term can dominate for large separation over the contribution from the potential
which behaves as $1/d^3$. We wish to investigate the effect of
this kinetic term on nucleon-nucleon scattering. There are two questions to
address: firstly how to properly formulate the inclusion of the term within the
low energy, semi-classical framework and secondly, how to characterize its
effects on the scattering. We will work with the hypothesis that the potential
term is in fact negligible, we discuss in the conclusion the actual physical
realization of this eventuality. With this hypothesis, the scattering
corresponds to geodesic motion in the induced metric on the sub-manifold.
The effect of this induced metric is to modify the equations of motion in a
rather complicated way. The utility of the differential geometric formulation
is less evident with respect to the corresponding work on scattering of BPS
monopoles\cite{9} or self dual vortices\cite{10}, primarily because of the lack
of symmetries and useful algebraic structures. Instead the equations are
better formulated as one complicated, exact, dynamical equation for the
coordinate corresponding to the relative separation, augmented by conservation
laws, which are now violated by perturbative corrections coming from the new
interaction, for the generators of rotation and for isorotation of the free
system. We make a further approximation, using the method of ``variation of
constants''\cite{11} to compute the scattering. The method essentially consists
of using the free trajectories in the interaction terms to produce evolution
of the previously conserved (constant) quantities. We find non-trivial
scattering for all initial polarizations.

The extraction of the implied nucleon-nucleon scattering is non-standard. Up to
now, most investigations have focussed directly on the projection onto
asymptotic, quantum nucleonic states. This requires a consistent interpretation
of the kinetic energy as supplying an induced metric on the sub-manifold. This
metric modifies the generally covariant Laplacian defined on the
sub-manifold which then should be included in the relevant Schr\"odinger
operator\cite{7}. We find that this procedure has not been fully
implemented\cite{5,12,13}. It remains to be seen whether directly quantizing
the
truncated system is phenomenologically viable. Actually we find this procedure
somewhat orthogonal to the incorporation of classical solitons within the
quantum theory. This incorporation is necessarily formulated in terms of the
semi-classical perturbative expansion, indeed the soliton is a {\it classical}
solution. In this article we give the construction of the semi-classical
nucleonic states using the Bohr-Sommerfeld (WKB) type quantization rules. These
states are then amenable to calculation of the scattering within the
semi-classical approximation.

In section 2 we give the equations of motion and the approximation methods
used.
In section 3 we give the analysis of the semi-classical quantum nucleon states.
In section 4 we summarize our results for the scattering and we terminate with
our conclusions in section 5.

\section{Equations of Motion and Approximation Methods}

In this section we will indicate briefly the path to deriving the equations of
motion and the approximations that we use in our treatement.
For a more detailed description we refer the reader to our previous article and
references therein\cite{8}.

With the definitions
\begin{equation}
{\cal L}^a_\mu(U)=-{i\over 2}\; tr[\tau^a \partial_\mu U U^\dagger]\label{pL}
\end{equation}
\begin{equation}
{\cal R}^a_\mu(U)=-{i\over 2}\; tr[\tau^a U^\dagger \partial_\mu U]\label{pR}
\end{equation}
\begin{equation}
D_{ab}(U) = {1\over 2}\;tr[\tau^a U \tau^b U^\dagger]\label{pD}
\end{equation}
where the field $U(\vec x,t)$ is an element of the $2\times 2$ matrix valued
representation of $SU(2)$, the Skyrme Lagrangean becomes
\begin{equation}
{\cal L}_{sk}={f_\pi^2\over 2} {\cal L}_\mu(U)\cdot{\cal L}^\mu(U)
- {1\over{4 e^2}}\biggl [{\cal L}_\mu(U)\cdot{\cal L}^\mu(U)
{\cal L}_\nu(U)\cdot{\cal L}^\nu(U) -
{\cal L}_\mu(U)\cdot{\cal L}^\nu(U) {\cal L}_\nu(U)\cdot{\cal
L}^\mu(U)\biggr].
\label{pLsk}
\end{equation}
We replace for $U(\vec x,t)$ in this Lagrangean with the product ansatz
\begin{equation}
U(\vec x,t) = A U_s(\vec x-\vec R_1) A^\dagger
	    B U_s(\vec x-\vec R_2) B^\dagger\label{pP}
\end{equation}
where $U_s(\vec x)=e^{i f(r) \hat r\cdot \vec \tau}$ is the single Skyrmion
with profile function $f(r)$, $A(t),B(t)\in SU(2)$ define the respective spin
and isospin orientations of the Skyrmions and $\vec R_1$ and $\vec R_2$ are
their
positions. This corresponds to the truncation\cite{7} to the relevant degrees
of freedom for the low energy behaviour. We obtain the kinetic energy by
keeping only the terms which contain time derivatives, these are automatically
quadratic, the remaining terms contain no time derivatives and comprise the
potential.

The product ansatz makes sense only for well separated Skyrmions, when they are
close, we expect strong deformations and, in fact, the minimal critical point
is conceived to be a toroidal configuration\cite{14}. A proposal
to globally, approximately describe the sub-manifold of low energy critical
points and the union of gradient flow curves joining these together has been
put forward by Atiyah and Manton\cite{15}. We do not consider, however, the
possibility of close encounters between the Skyrmions and hence the product
ansatz should yield at least the correct, leading asymptotic result.
The induced kinetic
energy has a well defined expansion in inverse powers of the separation
$d$. The zero order terms correspond to the free kinetic energy of
each individual Skyrmion. The leading correction behaves as $1/d$ and
contributes to a spin-spin and tensor interaction between the Skyrmions, as
we\cite{8} and Schroers\cite{16} have found. Thus the kinetic energy is given
by
\begin{equation}
\begin{array} {l}
T = {1\over 4} M \dot{\vec d^{\;2}} + 2 \Lambda \bigl({\cal L}^a(A)\, {\cal
L}^a(A) + {\cal L}^a(B)\,{\cal L}^a(B)\bigr)
\\
\qquad\qquad\qquad+ {\Delta\over d} \epsilon^{iac}\epsilon^{jbd}\;{\cal R}^c(A)
\,{\cal R}^d(B)\;
\bigl(\delta^{ij}-\hat{d}^i \hat{d}^j\bigr)\, D_{ab}(A^\dagger B) +O(1/d^2)
\end{array}\label{pT}
\end{equation}
where $M$ is the Skyrmion mass, $\Lambda$ its moment of inertia, $\Delta=2 \pi
\kappa^2 f_\pi^2$, $F(r)\sim \kappa/r^2$ at large
$r$, ${\cal L}^a (A)\equiv {\cal L}^a_0(A)$, ${\cal R}^a (A)\equiv
{\cal R}^a_0(A)$ and $\hat d = \vec d/d=(\vec
R_1-\vec R_2)/d$. The
potential has been calculated by many authors\cite{4,5,17,22}. It has a leading
term which contributes to the tensor channel and behaves as $1/d^3$, apart from
the ``free'' contribution which gives the mass of each Skyrmion:
\begin{equation}
V = 2 M + 4\pi f_\pi^2 \kappa^2 {(1-\cos\theta) (3(\hat n\cdot\hat
d)^2-1)\over d^3}.\label{pV}
\end{equation}
The complete Lagrangean of the system is given by
\begin{equation}
L=T-V.\label{pGL}
\end{equation}
In this paper we focus on the effects of the kinetic term. The equations of
motion arising from the kinetic term are neither illuminating nor practical. It
is useful just to write down the approximate equations. The free Lagrangean is
completely integrable and corresponds exactly to two free, spherically
symmetric tops. We can express the approximate equations as corrections to
conservation laws using the approximation of Lagrange\cite{11}.
Here we take the Poisson brakets implied by the free dynamics to compute the
corrections implied by the interaction to the equations of motion.
This yields a practical set of
equations which must be treated, however, with further perturbation measures
\begin{equation}
\begin{array} {l}
{d\over dt}\dot d^k = -{2\Delta\over M d^2}\biggr[\delta^{ij}\hat d^k +
\delta^{jk}\hat
d^i +\delta^{ik}\hat d^j - 3\hat d^i \hat d^j \hat d^k\biggl] \epsilon^{iac}
\epsilon^{jbd} {\cal R}^c(A) {\cal R}^d(B) D_{ab}(A^\dagger B)
\\
{d\over dt}{\cal L}^k(A) =  {\Delta\over 2 M d} \epsilon^{iac} \epsilon^{jbd}
{\cal R}^c(A) {\cal R}^d(B) \bigl(\delta^{ij}-\hat d^i\hat d^j\bigr)
\epsilon^{kef} D_{fa}(A)D_{eb}(B)
\\
{d\over dt}{\cal L}^k(B) =  {\Delta\over 2 M d} \epsilon^{iac} \epsilon^{jbd}
{\cal R}^c(A) {\cal R}^d(B) \bigl(\delta^{ij}-\hat d^i\hat d^j\bigr)
\epsilon^{kef} D_{ae}(A^\dagger )D_{fb}(B)
\\
{d\over dt}{\cal R}^k(A) = - {\Delta\over 2 M d}\epsilon^{iac} \epsilon^{jbd}
{\cal R}^d(B)\bigl(\delta^{ij}-\hat d^i\hat d^j\bigr)
  \biggl[\epsilon^{kcf} {\cal R}^f(A) D_{ab}(A^\dagger B) +
\epsilon^{kaf} D_{fb}(A^\dagger B) {\cal R}^c(A)\biggr]
\\
{d\over dt}{\cal R}^k(B) = - {\Delta\over 2 M d}\epsilon^{iac} \epsilon^{jbd}
{\cal R}^c(A)\bigl(\delta^{ij}-\hat d^i\hat d^j\bigr)
  \biggl[\epsilon^{kdf} {\cal R}^f(B) D_{ab}(A^\dagger B) +
\epsilon^{kbf} D_{af}(A^\dagger B) {\cal R}^d(B)\biggr].
\end{array}\label{pEM}
\end{equation}
The method of variation of constants is ideal for the following analysis. Here
we replace the variables in the right hand side of the ``conservation
equations'' by their trajectories coming from the free dynamics.
This gives rise to a ``variation'' of the previously conserved
``constants''. The procedure can be iterated to give higher order corrections.
One should however always maintain consistency with the first (Lagrange)
approximation. In fact, the method of ``variation of constants'' is
only useful with respect
to the equation for $\vec d$, where it gives the scattering trajectory. The
change in the spin or isospin governed by the subsequent four equations cannot
be used in conjunction with this approximation method because of the long range
of the interaction. The results give an infinite change in these angular
momenta, which is not reliable.

\section{Semiclassical Nucleon states}

In this section we will construct quantum states corresponding to the nucleons,
using the semi-classical Bohr-Sommerfeld quantization rules applied to the
classical motion of the free Skyrmion. To apply these rules, we must find the
classical action-angle variables of the system and impose quantization
conditions\cite{18,19}.
This enables us to treat a Skyrmion as a nucleon consistently within the
context of the semi-classical expansion. Going directly to the quantization of
the Skyrmion\cite{3,4,5,22} is perhaps not in the spirit of the semi-classical
formulation which is, we believe, incumbent upon us once we have begun to
consider classical solutions. In addition, we find that with the proper
semi-classical
quantization we can make physical sense of the specific, classical Skyrmion
scattering trajectories for the scattering of nucleons.
We start
with the Lagrangean for a single Skyrmion in the center of mass frame which is
given by:
\begin{equation}
\begin{array} {l}
L = - M +  \lambda \,tr\bigl[ \dot A^\dagger \dot A\bigr]
\\
\qquad = - M +  2 \lambda \,\sum_{i=0}^{3} \dot a_i^2
\end{array}\label{pL1}
\end{equation}
where $\dot a_i$ is the time derivative of the $a_i$ from the parametrization
$A=a_0+i\vec a\cdot \vec\tau$ with $a_0^2 + {\vec a}^2=1$ of the matrix
defining the rotational
characteristics of the Skyrmion. Following the treatment of the motion of a
rigid body in space, we go to the system of coordinates corresponding to the
usual Euler angles:
\begin{equation}
A = e^{-i\alpha \tau_3/2}\,e^{-i\beta \tau_2/2}\,e^{-i\gamma
\tau_3/2}\label{pA}
\end{equation}
where $\alpha,\gamma\in [0,2\pi]$ and $\beta\in [0,\pi]$. The $a_i$ are related
to the Euler angles by the equations:
\begin{equation}
\begin{array} {l}
a_0 = \cos({\beta\over 2})\cos({\alpha+\gamma\over 2})
\\
a_1 = \sin({\beta\over 2})\sin({\alpha-\gamma\over 2})
\\
a_2 = - \sin({\beta\over 2}) \cos ({\alpha-\gamma\over 2})
\\
a_3 = - \cos({\beta\over 2}) \sin({\alpha+\gamma\over 2}).
\end{array}\label{pAE}
\end{equation}
Replacing them into the Lagrangean (\ref{pL1}) we get
\begin{equation}
L = - M + {1\over 2}\lambda\,\bigl[ \dot \alpha^2+ \dot \beta^2 +\dot \gamma^2
+ 2 \dot \alpha \dot \gamma \cos \beta\bigr]\label{pLA}
\end{equation}
which is the well known Lagrangean for rigid body motion.
In this case, the Euler angles are used to describe the motion of the body
fixed axes relative to the laboratory fixed axes. In our case, the
identification of the Euler angles will be completely different.

The action-angle variables $J_i$ are defined by\cite{20}
\begin{equation}
J_i = \oint p_i \,dq_i\label{pAV}
\end{equation}
where $p_i$ is the momentum conjugate to the coordinate $q_i$, and the
integral is taken along a closed path followed by the system during one
period in the plane $(q_i,p_i)$ of phase space. In our case there are three
such variables, one for each angle. Due to the cyclic nature of the angles
$\alpha$ and $\gamma$, $J_\alpha$ and $J_\gamma$ are readily computed:
\begin{equation}
J_\alpha = \oint p_\alpha dq_\alpha= 2\pi p_\alpha=
2 \pi\lambda\bigl[ \dot \alpha + \dot \gamma \cos \beta\bigr]\label{pAa}
\end{equation}
\begin{equation}
J_\gamma = \oint p_\gamma dq_\gamma= 2\pi p_\gamma=
2\pi\lambda\bigl[ \dot \gamma + \dot \alpha \cos \beta\bigr].\label{pAg}
\end{equation}
In contrast, $J_\beta$ is much more complicated for general orbits, however,
for the type of motions that we will consider it is zero. The motion of a free
Skyrmion corresponds simply to uniform isorotation about a fixed axis. The
equations of motion of the Lagrangean (\ref{pLA}) are
\begin{equation}
\begin{array} {l}
\ddot\alpha + \ddot\gamma\cos\beta-\dot\gamma\dot\beta\sin\beta=0
\\
\ddot\beta+ 2\dot\alpha\gamma=0
\\
\ddot\gamma + \ddot\alpha\cos\beta-\dot\alpha\dot\beta\sin\beta=0
\\
\end{array}
\end{equation}\label{eqangles}
Without loss of generality we choose the axis of isorotation to be the third
axis in isospace. Then the time evolution is clearly given by linear time
dependence of $\alpha$ and/or $\gamma$ with $\beta=0$ or $\pi$.

The Bohr-Sommerfeld quantization condition, {\it derived} by path integral
methods\cite{18,19} is
\begin{equation}
W = \sum_i J_i = (n+\xi) h, \qquad n= 0,\pm 1,\pm 2,\cdots .
\end{equation}\label{pW}
$\xi$ is a correction factor arising from the functional integral over Gaussian
fluctuations about the classical trajectory. In zero order\cite{18} we can
neglect $\xi$. Then we find
\begin{equation}
J_\alpha+ J_\gamma = n h, \qquad n= 0,\pm 1,\pm 2,\cdots
\end{equation}
in perfect accord with the original Bohr-Sommerfeld quantization rules.
In order to precisely fix the time dependence of $\alpha$ and $\gamma$
corresponding to nucleon states we must make reference to the isospin and spin
generators. These are given by, respectively:
\begin{equation}
\begin{array} {l}
\hat{\cal L}^3 = 2 \lambda{\cal L}^3= -{J_\alpha\over 2\pi}
\\
\hat{\cal R}^3 = 2 \lambda{\cal R}^3= -{J_\gamma\over 2\pi}
\end{array}\label{pLhatRhat}
\end{equation}
and satisfy the usual angular momentum algebra. The symmetries of the Skyrmion
dictate that spin is equal to isospin, thus
\begin{equation}
|J_\alpha| = |J_\gamma|.
\end{equation}
Then applying the quantization condition with $n=1$, gives
\begin{equation}
|\hat{\cal L}^3| = |\hat{\cal R}^3| = {1\over 2}.
\end{equation}
We will follow the convention of Adkins {\it et al}~~\cite{3} who define the
isospin generator as
\begin{equation}
I_3 = - \hat{\cal L}^3\label{ppL}
\end{equation}
and the spin generator as
\begin{equation}
J_3 = \hat{\cal R}^3.\label{ppR}
\end{equation}
Consider first the case $\beta=0$. Replacing this in equation (\ref{pAa}),
(\ref{pAg}), (\ref{pLhatRhat}), (\ref{ppL}) and (\ref{ppR}) we get
\begin{equation}
\begin{array} {l}
I_3=\lambda(\dot\alpha+\dot\gamma)\equiv 2 \lambda \omega=\pm{1/2}
\\
J_3=-\lambda(\dot\alpha+\dot\gamma)\equiv -2 \lambda \omega=\mp{1/2}
\end{array}\label{pb0}
\end{equation}
with $(\alpha+\gamma)/2=\phi(t) = \omega t + \phi_0$.
This type of angular motion produces Skyrmions with spin and isospin
antiparallel, and eventually the states $|p\downarrow>$ and $|n\uparrow>$.
The proton state corresponds to isospin $+1/2$ along the 3 axis in isospace
while the neutron corresponds to isospin $-1/2$. The selection of the third
axis, we emphasize, is purely convention. States polarized along some other
axis in isospace simply corresponds to linear superpositions of our states.
The corresponding beams would be a mixture of protons and neutrons.

Replacing this solution in equation (\ref{pAE}) gives
\begin{equation}
\begin{array} {l}
A = \cos\phi(t) - i \sin\phi(t) \tau^3
\\
\quad = e^{-i \phi(t)\tau^3/2}
\end{array}\label{pAA1}
\end{equation}
where $\omega>0$ corresponds to the state$|p\downarrow>$ and $\omega<0$ to
$|n\uparrow>$.

For $\beta=\pi$
\begin{equation}
\begin{array} {l}
I_3=\lambda(\dot\alpha-\dot\gamma)\equiv 2 \lambda \dot\omega=\pm 1/2
\\
J_3=-\lambda(\dot\gamma-\dot\alpha)\equiv 2 \lambda \dot\omega=\pm 1/2
\end{array}\label{pbpi}
\end{equation}
with $(\alpha-\gamma)/2=\psi(t) = \omega t + \psi_0$ and the corresponding
matrix
\begin{equation}
\begin{array} {l}
A = i\bigl[ \sin\psi(t) \tau^1-\cos\psi(t)\tau^2\bigr]
\\
\quad = -i e^{-i \psi(t)\tau^3/2}\,\tau^2\, e^{i \psi(t)\tau^3/2}
\end{array}\label{pAA2}
\end{equation}
represents the state $|p\uparrow>$ with $\omega>0$ and $|n\downarrow>$ with
$\omega<0$. We can see the similarity of the forms for the matrices $A$
corresponding to nucleon states and the wave functions obtained in the directly
quantum version of Adkins {\it et al}~~\cite{3}. In the next section we study
the scattering of the nucleons as described by these states, using the
geodesics
approximation.

\section{Nucleon-Nucleon scattering}

The nucleon states found in the previous section are ideally suited for our
treatment of nucleon-nucleon scattering using the approximation method of the
``variation of constants''. We simply replace into the right hand side of
equation (\ref{pEM}) the quantized classical trajectories found in the
previous section.
To calculate the change induced in the previously constant quantities we
integrate the equations from $t=-\infty$ to $t=+\infty$ over one free
trajectory.  This is useful for the equation governing the relative motion
(\ref{pEM}). For the angular variables the perturbation method is too crude.
It is perhaps a reasonable perturbative scheme locally in time, however, to
integrate the motion
over all time leads to infinite changes in the various angular momenta. We do
not trust such results and we suggest that the approximation should be iterated
several times to obtain a better approximation to the local (in time)
perturbation of the free trajectories. This could then perhaps be integrated
over all time. At the root of the divergence is the long range ($1/d$) nature
of the interaction term. Clearly for a physical theory with non-zero pion mass,
this interaction would be cut-off by the usual Yukawa factor, rendering all
such integrations finite. Thus we will
consider the variation of the angular momenta only in so far as making the
observation that the equations of motion with the interaction imply non-trivial
spin flip and charge flip scattering. These effects are compatible with the
exchange of charge carrying (pions) and spin carrying (vector mesons)
intermediate particles. The first equation of (\ref{pEM}) which describes
the relative position is of course sufficient to obtain the exclusive
scattering. We present below the results for the scattering of specially
polarized nucleons, described in our semi-classical formalism.

It is interesting to
observe that in our formalism an additional parameter arises which describes
the
initial state of two incoming, polarized nucleons. This parameter, along with
the impact parameter, the initial velocity and the direction of polarization,
actually selects the particular
scattering trajectory followed by the nucleons. The parameter describes the
relative orientation of the Skyrmions at a fixed time. It plays in fact a role
similar to a hidden variable. An incoming pair of physical nucleons, in our
formalism, has a fixed value for this parameter, which is only ``measured''
after
the scattering takes place. In a physical experiment consisting of incoming
beams of nucleons giving rise to collisions or scattering of pairs of nucleons,
the value of this parameter will be uniformly distributed. A similar parameter
arises in the case of the scattering of BPS monopoles\cite{21}.

There is an immediate separation of the scattering into two cases, depending on
whether $D_{ab}(A^\dagger B)$ is time independent or not. When it depends on
time for large values of the ratio $(\omega\gamma/v)$ there is an exponential
suppression
of the scattering where $v$ is the relative velocity and $\gamma$ the impact
parameter. This is quite evident, for slowly translating Skyrmions, the
prescribed rotations imposed by selecting semi-classically quantized nucleon
states have the effect of averaging the interaction to zero.

We find the following expression for time variations of the previously
constant relative momentum $\vec p = (M/2) \dot{\vec d}$:
\begin{center}
{\bf $D_{ab}(A^\dagger B)$ time independent:}
\end{center}
\begin{equation}
\begin{array} {l}
p\uparrow p\uparrow
\\
n\downarrow n\downarrow
\\
p\downarrow p\downarrow
\\
n\uparrow n\uparrow
\end{array}
\qquad {d\over dt} p^k = - {\Delta\omega^2\over d^2} \cos({2\delta}) \hat
d^k\label{p1}
\end{equation}

\begin{equation}
p\uparrow p\downarrow
\qquad
\begin{array}{l}
{d\over dt} p^k = - {\Delta\omega^2\over d^2}\bigl[
\hat d^k + 4 r^k \hat r\cdot\hat d - 6 \hat d^k (\hat r\cdot\hat d)^2 \bigr]
\\
\qquad\qquad\qquad\qquad\hat r^k = (-\sin(\delta),\cos(\delta),0)
\end{array}\label{p2}
\end{equation}

\begin{equation}
n\uparrow n\downarrow
\qquad
\begin{array}{l}
{d\over dt} p^k = - {\Delta\omega^2\over d^2}\bigl[
\hat d^k + 4 r^k \hat r\cdot\hat d - 6 \hat d^k (\hat r\cdot\hat d)^2 \bigr]
\\
\qquad\qquad\qquad\qquad\hat r^k = (\sin(\delta),-\cos(\delta),0)
\end{array}\label{p3}
\end{equation}

\begin{center}
{\bf $D_{ab}(A^\dagger B)$ time dependent:}
\end{center}
\begin{equation}
\begin{array} {l}
p\uparrow n\downarrow
\\
p\downarrow n\uparrow
\end{array}
\qquad {d\over dt} p^k = {\Delta\omega^2\over d^2} \cos(4\omega t+\epsilon)\hat
d^k\label{p4}
\end{equation}

\begin{equation}
p\uparrow n\uparrow
\qquad
\begin{array}{l}
{d\over dt} p^k = {\Delta\omega^2\over d^2}\bigl[
\hat d^k + 4 r^k \hat r\cdot\hat d - 6 \hat d^k (\hat r\cdot\hat d)^2 \bigr]
\\
\qquad\qquad\qquad\qquad\hat r^k = (-\sin(2\omega t + \epsilon),
\cos(2\omega t + \epsilon),0)
\end{array}\label{p5}
\end{equation}

\begin{equation}
p\downarrow n\downarrow
\qquad
\begin{array}{l}
{d\over dt} p^k = {\Delta\omega^2\over d^2}\bigl[
\hat d^k + 4 r^k \hat r\cdot\hat d - 6 \hat d^k (\hat r\cdot\hat d)^2 \bigr]
\\
\qquad\qquad\qquad\qquad\hat r^k = (-\sin(2\omega t + \epsilon),
-\cos(2\omega t + \epsilon),0)
\end{array}\label{p6}
\end{equation}
with $\vec d(t)$ in the right hand side given by
\begin{equation}
(d_x,d_y,d_z) = (v t ,\gamma,0).
\end{equation}
$A(t)$ and
$B(t)$ are given by the angular dependence found in the previous section
corresponding to the associated nucleon states, with $\delta=\phi_0^A-\phi_0^B$
and $\epsilon=\phi_0^A+\phi_0^B$ in self evident notation. The right hand
sides can be interpreted, at this level of our approximation, as coming from
a spin-spin channel and a tensor channel interaction. We stress that this is
only a correspondance, the true effect of the kinetic term is to supply a
non-trivial connection in the geodesic equations on the low energy
sub-manifold and not to modify the potential.

To find the actual change in $\vec p$ and
hence the scattering angle, we integrate these equations from $t=-\infty$ to
$t=+\infty$. The expressions for the scattering of $p$ on $n$ each contain a
time dependent $A^\dagger B$. When integrated these yield an exponentially
suppressed variation in the dimensionless group $(\omega\gamma/v)$
\begin{equation}
\sim e^{-({\omega\gamma\over v})}.
\end{equation}
Thus in the limit $v\to 0$ we get negligible scattering in these cases. For the
cases $pp$ or $nn$ we find the scattering angle depends on a variable $\delta$
which corresponds to the phase lag between the rotation of $A(t)$ and $B(t)$.

We give the $p\uparrow p\uparrow$ scattering in detail to make things
concrete. Here we find
\begin{equation}
\begin{array} {l}
p^k(t) =-\Delta\,\omega^2 \cos(2\delta) \int_{-\infty}^t {v^k t + \gamma^k
\over
(v^2 t^2 + \gamma^2)^{3/2}}dt + {M\over 2} v^k
\\
\qquad\qquad= -\Delta\,\omega^2 \cos(2\delta)\Bigl( -{v^k \over v^2 (v^2 t^2 +
\gamma^2)^{1/2}} + {\gamma^k t\over \gamma^2 (v^2 t^2 + \gamma^2)^{1/2}}\Bigr)
+ {M\over 2} v^k.
\end{array}\label{intex}
\end{equation}
This yields
\begin{equation}
p^k(+\infty) = -2 \Delta\,\omega^2 \cos 2\delta {\gamma^k\over \gamma^2 v} +
{M\over 2} v^k
\end{equation}
from which we calculate the cosine of the scattering angle
\begin{equation}
\begin{array}{l}
\cos\theta={\vec p(+\infty)\cdot\vec p(-\infty)\over |\vec p(+\infty)| |\vec
p(-\infty)|}
\\
\qquad\qquad= {M\gamma v^2\over 4 \Delta} {1\over\bigl( {M^2\gamma^2 v^4\over
16
\Delta^2} + \omega^4 \cos^2 2\delta \bigr)^{1/2}}.
\end{array}\label{angle}
\end{equation}
This is the first analytical calculation of nucleon-nucleon scattering from
essentially first principles, without recourse to {\it ad hoc} models or
potentials. We reemphasize that the Skyrme model is in principle derivable from
QCD and $f_\pi$ and $e$ are, as such, calculable parameters and, in that sense
this is also a QCD calculation.

We wish to point out that in the limit that the initial velocity vanishes,
for fixed $\omega$ and $\gamma$, we recover 90$^\circ$ scattering. This is,
however, not so surprising as it is a property also shared by the Coulomb and
many other
interactions treated with our approximation. 90$^\circ$ scattering is hardly
remarkable except at zero impact parameter, where of course, it is impossible
to avoid the region of close proximity of the nucleons and it seems important
that the configurations pass through the minimal, toroidal configuration.

We have up to now considered scattering in the $(x,y)$ plane with spin
polarized in the orthogonal $z$ direction. The tensorial nature of the
interacton implies that the forces depend on the angle between the axis of
separation and spin polarization. If we choose the spin polarization along an
axis tilted with respect to the normal to the initial scattering plane we get
complicated, three dimensional scattering trajectories.

\section{Conclusions}

In this paper we have taken to its logical conclusion the study of
nucleon-nucleon
scattering in the Skyrme model using the method of truncation to the relevant
finite
number of degrees of freedom and their ensuing dynamics as implied by an
induced metric and potential. We have restricted ourselves to large separations
between the Skyrmions. In this regime, the Skyrmions are well described to
leading order by the product ansatz and the manifold of collective
coordinates
is parametrized by the variables of the product ansatz. The induced metric,
calculated to leading order behaves as $1/d$ while the induced potential
behaves as $1/d^3$. In principle there can be a region where the
induced metric contribution dominates and we can neglect the potential. It is
exactly this point of view which we have taken to its logical conclusion. We
find that the metric induces an interaction which can be interpreted, within
our approximation method, as a spin-spin and a tensor interaction.
Unfortunately
it seems that the domination by the metric term is not physically realized. The
induced kinetic term is multiplied by essentially the frequencies of angular
rotation of the Skyrmions while the potential term has two extra powers of the
separation in the denominator. Thus for the kinetic term to dominate, the
frequencies should be much larger than the separation, in units where distance,
time and energies are all measured in MeV. The frequencies are fixed by the
Bohr-Sommerfeld quantization rules which give
\begin{equation}
\omega \sim 100 MeV.
\end{equation}
This corresponds to a region of validity for a separation of about 3 fm and
greater. However there is much latitude
available since the values of $f_\pi$ and $e$ which go into determining
$\omega$ are fixed only by choosing two experimental inputs. $f_\pi$, $e$ can
vary as much as 10-30\% thus we do not feel overly concerned with exceeding the
regime of validity. Our approximation would of course be better justified for
the case of $\Delta\!-\!\Delta$ scattering where $\omega\sim 300$MeV
corresponding
to a separation of 1 fm. In any case, we do believe however, that it is not
physically reasonable to consider the scattering of nucleons with the metric
term alone and we expect a contribution from the potential term which is of
the same order of magnitude.  We do not expect, however, any great,
qualitative modification of the scattering upon inclusion of the potential
term, it is of similar strength but actually contributes only in the tensor
channel for the case of massless pions.

We find our work as a prescriptive and requisite endeavour to investigate
the nucleon-nucleon interaction arising from the kinetic term. We have shown
how to properly formulate the nucleon states within the semi-classical
approximation. We have treated the scattering and computed the scattering
angles in a systematic perturbative method. Future work should include
consideration of a non-zero pion mass, which leads to a central channel
interaction, a better control of the perturbation method,a departure from the
product ansatz and a proper treatment of the region of close proximity, to
test the validity of our formalism in the phenomenology of low energy
nucleon-nucleon scattering and of the static quantum states in this sector.

\acknowledgments
We thank D. Caenepeel, J. LeTourneux, R. Mackenzie, J. M. Pearson and M.
Temple-Raston for useful discussions. This work supported in part by NSERC of
Canada and FCAR of Qu\'ebec.

\end{document}